\documentclass[reprint,showpacs,noeprint,amsmath,a4paper,amssymb,xcolor=dvipsnames]{revtex4-1}
\pdfoutput=1

\usepackage[pdftex]{graphicx}
\usepackage{graphicx}
\usepackage{dcolumn}
\usepackage{bm}
\usepackage{datetime}
\usepackage{float}
\usepackage[pdftex,colorlinks=true,urlcolor=black,linkcolor=black]{hyperref} 
\usepackage{microtype}
\usepackage{epstopdf} 
\usepackage{amsmath}
\usepackage{amssymb}
\usepackage{braket}
\usepackage{xcolor}
\usepackage[version-1-compatibility,binary-units,input-product=x,abbreviations=true]{siunitx}
\include{mathdefs.tex}
\newcommand{\diff}{\text{d}}

\newcommand{\Note}[1]{\textcolor{black}{#1}}
\newcommand{\NoteN}[1]{\textcolor{black}{#1}} 

\newcommand{\sfigl}[1]{\textbf{#1}}
\newcommand{\figfirst}[1]{#1}

\DeclareMathOperator{\Tr}{Tr}

\renewcommand{\vec}[1]{\mathbf{#1}}

\newcommand{\vac}{\left| 0 \right\rangle}

\newcommand{\cupleft}{c^\dagger_{\uparrow \textrm{L}}}
\newcommand{\cupright}{c^\dagger_{\uparrow \textrm{R}}}
\newcommand{\cdownleft}{c^\dagger_{\downarrow \textrm{L}}}
\newcommand{\cdownright}{c^\dagger_{\downarrow \textrm{R}}}

\newcommand{\unsim}{\mathord{\sim}} 
\def\abs#1{\left| #1 \right|}

\newcommand{\beq}{\begin{equation}}
\newcommand{\eeq}{\end{equation}}
\newcommand{\beqnarray}{\begin{eqnarray}}
\newcommand{\eeqnarray}{\end{eqnarray}}

\renewcommand{\figurename}{FIG}

\makeatletter
\newcommand*{\balancecolsandclearpage}{%
  \close@column@grid
  \cleardoublepage
  \twocolumngrid
}
\makeatother

\begin{document}

\title{Experimental Characterization of Two-Particle Entanglement through Position and Momentum Correlations}

\author{Andrea Bergschneider}
	\thanks{Present address: Institut f\"ur Quantenelektronik, Auguste-Piccard-Hof 1, 8093 Z\"urich, Switzerland}
	\affiliation{Physikalisches Institut der Universit\"at Heidelberg, Im Neuenheimer Feld 226, 69120 Heidelberg, Germany}

\author{Vincent M. Klinkhamer}
	\affiliation{Physikalisches Institut der Universit\"at Heidelberg, Im Neuenheimer Feld 226, 69120 Heidelberg, Germany}
\author{Jan~Hendrik~Becher}
	\affiliation{Physikalisches Institut der Universit\"at Heidelberg, Im Neuenheimer Feld 226, 69120 Heidelberg, Germany}
\author{Ralf~Klemt}
	\affiliation{Physikalisches Institut der Universit\"at Heidelberg, Im Neuenheimer Feld 226, 69120 Heidelberg, Germany}
\author{Lukas~Palm}
	\affiliation{Physikalisches Institut der Universit\"at Heidelberg, Im Neuenheimer Feld 226, 69120 Heidelberg, Germany}
\author{Gerhard~Z\"urn}
	\affiliation{Physikalisches Institut der Universit\"at Heidelberg, Im Neuenheimer Feld 226, 69120 Heidelberg, Germany}
\author{Selim~Jochim}
	\affiliation{Physikalisches Institut der Universit\"at Heidelberg, Im Neuenheimer Feld 226, 69120 Heidelberg, Germany}
\author{Philipp~M.~Preiss}
	\email{preiss@physi.uni-heidelberg.de}
	\affiliation{Physikalisches Institut der Universit\"at Heidelberg, Im Neuenheimer Feld 226, 69120 Heidelberg, Germany}

\maketitle

\textbf{Quantum simulation is a rapidly advancing tool to gain insight into complex quantum states and their dynamics. Trapped ion systems have pioneered deterministic state preparation and comprehensive state characterization, operating on localized and thus distinguishable particles \cite{Blatt2012}. With ultracold atom experiments, one can prepare large samples of delocalized particles, but the same level of characterization has not yet been achieved \cite{Gross2017}. Here, we present a method to measure the positions and momenta of individual particles to obtain correlations and coherences. We demonstrate this with deterministically prepared samples of two interacting ultracold fermions in a coupled double well \cite{Murmann2015}. As a first application, we use our technique to certify and quantify different types of entanglement \cite{Ghirardi2002,Dowling2006,Bonneau2018}.}

\begin{figure}
        \centering
				\includegraphics[width=\columnwidth]{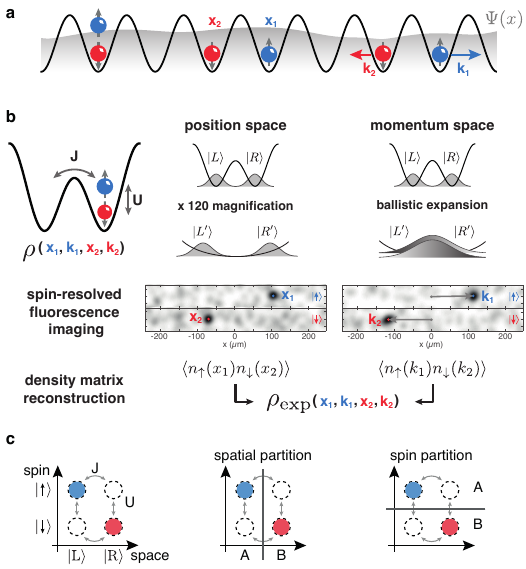}
        \caption{Detection of many-body systems in conjugate bases. a. A quantum system is defined by its many-body wavefunction $\Psi(\vec{x})$ (gray). \Note{A system with number fluctuations} can be characterized naturally with correlation functions in the positions $x_i$ or the momenta $k_i$ of its constituent particles. b. We measure single-particle spin-resolved correlations in position and momentum space for a two-site Fermi-Hubbard system, which we use to infer information about the density matrix $\rho$ of the initial state. c. The spinful Hubbard dimer forms a four-mode system. Different types of quantum correlations, or entanglement, can emerge between subsystems defined by spatial or spin partitions (A and B).}
        \label{fig:first}
\end{figure}

Ultracold atoms in optical lattice systems can be probed through site-resolved imaging \cite{Ott2016}. This procedure gives access to correlations on the level of individual particles and directly reveals charge and spin order in lattice systems \cite{Gross2017}. Many important aspects of the available quantum states, however, are not accessible by position-space imaging alone: Properties such as long-range coherence, currents and phase fluctuations are related to off-diagonal order, or coherences, in the many-body states. Time-of-flight imaging of quantum gases in momentum space can in principle probe such coherences \cite{Folling2005}, but has only been possible for systems of many particles, often leading to inhomogeneous averaging. A central goal for ultracold atom experiments is the development of new methods that access real-space order as well as coherences (Fig.~\ref{fig:first}a). 

In this letter, we show that single-particle resolved measurements of atomic momenta can reveal the coherence properties of small systems \cite{Bonneau2018,Yannouleas2018}. We demonstrate this for a Fermi-Hubbard double well \cite{Murmann2015,Desbuquois2017}, for which we measure two-body correlations in position and momentum space (Fig.~\ref{fig:first}b). This approach enables us to tightly constrain the full density matrix of the two-particle system in different regimes of attractive, repulsive, or vanishing interaction. 

The key motivation for such a characterization is to describe a state through its entanglement properties \cite{Tichy2011}. The presence of entanglement certifies the non-separability of a state with respect to a particular partitioning of the Hilbert space and constitutes the most prominent difference between classical and quantum mechanics \cite{Amico2008,Horodecki2009}. The characterization of entanglement can be useful for, e.g., applications in quantum metrology \cite{Pezze2018}. 

In lattice systems of ultracold atoms, different types of entangled states may occur: For states with exactly one particle per site, entanglement between internal degrees of freedom of localized particles can be described by spin models similar to trapped ion or superconducting qubit systems. Such states can be created in optical lattices \cite{Dai2016}, or through deterministic or probabilistic schemes in optical tweezers \cite{Kaufman2015, Lester2018}. Conceptually more challenging situations occur for indistinguishable, mobile particles: In such cases, particles are not distinguishable by their spatial location and cannot be identified as the carriers of quantum correlations. It is then more appropriate to consider the mode entanglement between different spatial regions \cite{Horodecki2009, Tichy2011}, as measured in recent experiments \cite{Cramer2013, Islam2015}. How to fully describe entanglement in situations with fluctuating local particle number and symmetrization constraints enforced by quantum statistics has been the subject of an intense debate in the literature \cite{Ghirardi2002, Zanardi2002,Wiseman2003,Dowling2006, Fukuhara2015, Mazza}. 

The scenario which we consider here allows the study of various forms of entanglement in a single experiment (Fig.~\ref{fig:first}c): We populate two spatial modes with two particles that are distinguishable by their spin state \cite{Murmann2015}. From one point of view, entanglement between the particles is driven by interaction-induced correlations between their motional states, which we identify experimentally.  On the other hand, we can focus on the entanglement between the two spatial modes, which is largest in the non-interacting regime with maximal particle number fluctuation. We verify and quantify entanglement, both between the spin modes and the spatial modes, through measurements of the R\'{e}nyi entropy \cite{Horodecki2009, Islam2015} and characterize its dependence on the interactions in the system. 

The experimental system consists of two \textsuperscript{6}Li atoms confined to a double-well potential formed by optical tweezers with a waist of 1.15\,$\mu$m and a wavelength of $\lambda = 1064$\,nm  \cite{Murmann2015}. The partially overlapping optical tweezers are tunnel-coupled with a rate $J$, and on-site interactions $U$ between atoms in hyperfine states $\ket\uparrow = \ket{F=1/2, m_F=+1/2}$ and $\ket\downarrow = \ket{F=3/2, m_F=-3/2}$ are controlled via a magnetic Feshbach resonance. Together, this results in a Hubbard Hamiltonian 
\begin{equation}
H = - J \sum_{\sigma} \left( \hat{c}^\dagger_{\text{L}\sigma} \hat{c}^{\vphantom{\dagger}}_{\text{R}\sigma} + \hat{c}^\dagger_{\text{R}\sigma} \hat{c}^{\vphantom{\dagger}}_{\text{L}\sigma} \right) + U \sum_{j = \text{L},\text{R}} \hat{n}_{j\downarrow}\hat{n}_{j\uparrow}
\label{eq:DWHamiltonian}
\end{equation}
with spatial modes L and R, where $\hat{c}^{(\!\dagger\!)}_{i\sigma}$ is the fermionic annihilation (creation) operator of a particle with spin $\sigma$ on site $i$ and $\hat{n}_{j\sigma} = \hat{c}^\dagger_{j\sigma} \hat{c}^{\vphantom{\dagger}}_{j\sigma}$. As demonstrated previously \cite{Murmann2015}, we initialize the system with one particle per spin state near its ground state by adiabatically transferring two deterministically prepared particles from a single optical tweezer to a dual-tweezer configuration. We then adiabatically tune the tunneling rate $J$ and the on-site interaction $U$ via the depth of the optical tweezers and the magnetic field, respectively.

\begin{figure}
        \centering
        \includegraphics[width=\columnwidth]{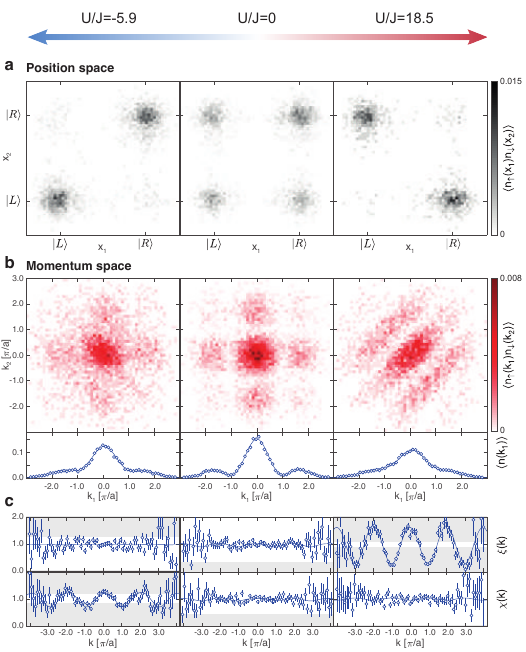}
				\caption{Correlations in the Hubbard dimer.
				a. The spin-resolved spatial correlation function $\langle n_\uparrow(\alpha) n_\downarrow(\beta)\rangle$ exhibits mostly double occupancies for attractive interactions (left), no correlations in the non-interacting case (center), and strong suppression of double occupancies for repulsive interactions (right).
				b. For non-interacting particles, the momentum-space correlation function $\langle n_\uparrow(k_1) n_\downarrow(k_2)\rangle$ (top row) is separable and shows an interference pattern in the single-particle coordinates. The single-particle coherence is visible as side peaks in the single-particle momentum density $\langle n(k_1) \rangle \equiv \langle n_\uparrow(k_1)\rangle +\langle n_\downarrow(k_1) \rangle$ (bottom row). For strong attractive (repulsive) interactions, single-particle coherence is suppressed, but interference patterns emerge along the diagonal (antidiagonal), which signals the presence of two-particle coherence. 
				c. Integrated momentum-difference and -sum correlations, expressed as pair correlators $\chi(k)$ and $\xi(k)$, respectively. Entanglement can be certified if the data extend into the gray-shaded regions. Error bars represent the standard error of the mean and continuous lines are obtained from reconstructed momentum space correlation functions (see Supplementary Information).}
        \label{fig:second}
\end{figure}

We now demonstrate how to obtain the position-space order and coherences of the experimentally initialized state. Our method is similar to recent proposals \cite{Bonneau2018,Yannouleas2018} and in close analogy to two-photon experiments working with near- and far-field correlations \cite{Taguchi2008}:  We combine atom-resolved measurements of correlations in position and momentum space, which we obtain through a novel, spin-resolved free-space detection method \cite{Bergschneider2018}. 

To access particle correlations in momentum space, we release the atoms from the tweezers into a large, elongated optical dipole trap. It allows expansion along the axis connecting the double well, while confining the atoms in the perpendicular directions (see Methods, \cite{Bergschneider2018}). After a ballistic expansion for one quarter trap period, the quantum state corresponds to the Fourier transform of the initial state. By using resonant single-atom imaging and separate exposures for the two spin states within each experimental realization (cf.\@ \cite{Bergschneider2018}), we record the particle momenta $k_1$ and $k_2$ (Fig.~\ref{fig:first}b). After several thousand iterations of the experiment, we can reconstruct the momentum correlation function $\langle n_\uparrow(k_1) n_\downarrow(k_2)\rangle$.

To probe the spatial correlations, we measure the occupation of each site in a spin-resolved manner. We make use of a position-mapping method, where we first project the quantum state on the single-site occupation basis by quickly decoupling the wells and then impart a site-specific momentum. This separates the spatial modes after time-of-flight for direct spatially resolved imaging (see Methods). We thus obtain the in-situ density distribution and determine spin-resolved correlation functions $\langle n_\uparrow(\alpha) n_\downarrow(\beta)\rangle$ (Fig.~\ref{fig:first}b), where $\alpha,\beta$ denote the spatial modes $\{\text{L},\text{R}\}$.

Figure~\ref{fig:second} shows the measured spin-resolved correlation functions for the Fermi-Hubbard dimer near its ground state for different interaction strengths (see Supplementary Information).
As theoretically expected, we directly observe that increasing repulsion (attraction) results in increasing anti-correlations (correlations) in position space (Fig.~\ref{fig:second}a). Simultaneously single-particle coherences disappear and a two-particle coherence appears, as visible in momentum space (Fig.~\ref{fig:second}b).
We analyze the data by extracting the pair correlators $\xi(d) = \frac{\int \diff \kappa \left\langle n_\uparrow(\kappa-d/2) n_\downarrow(\kappa+d/2) \right\rangle}{\int \diff \kappa \left\langle n_\uparrow(\kappa-d/2) \right\rangle \left \langle n_\downarrow(\kappa+d/2) \right\rangle}$ and $\chi(s) = \frac{\int \diff \kappa \left\langle n_\uparrow(\kappa+s/2) n_\downarrow(-\kappa+s/2) \right\rangle}{\int \diff \kappa \left\langle n_\uparrow(\kappa+s/2) \right\rangle \left \langle n_\downarrow(-\kappa+s/2) \right\rangle}$ in the relative and center-of-mass momentum coordinates $d = k_1 - k_2$ and $s = k_1 + k_2$, respectively, analogous to the noise correlation experiments performed in \cite{Folling2005}  (Fig.~\ref{fig:second}c). For all measurements with interactions, the observed signal differs from \num{1}, confirming the presence of pair correlations.

The observed correlations qualitatively agree with the expectations for the ground state in this highly controlled scenario.  An essential question is how to use such experimental data to certify and quantify entanglement \cite{Fukuhara2015, Kaufman2015,Dai2016,Lester2018,Cramer2013,Islam2015}. Here, we are specifically interested in entanglement between particles, which we treat as distinguishable via their spin. Qualitatively, this question can be addressed by an entanglement witness (see Supplementary Information). The witness probes incompatibility with general product states (grey regions in Fig.~\ref{fig:second}c) and from our data certifies entanglement between particles for $\left|U/J\right| \gtrapprox 5$.

\begin{figure}[!tb]
        \centering
				\includegraphics[width=\columnwidth]{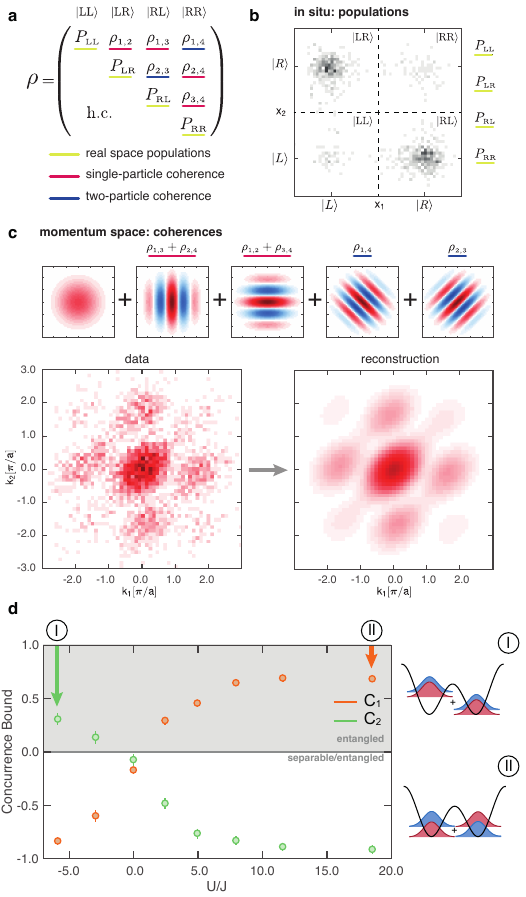}
        \caption{Evaluation of the concurrence from the measured correlation functions. a. The dimer density matrix $\rho$ contains the full state information of real-space order as well as coherence properties. b. We obtain the \textit{in-situ} populations directly from the position-space correlation function. c. The magnitude and phase of the single- and two-particle coherences are encoded in the momentum-space correlation function as oscillations along the one- and two-particle coordinates. d. The lower bounds $C_1,C_2$ of the concurrence certify entanglement (gray region) for all interacting systems studied in our experiments, with error bars corresponding to one standard deviation of statistical and systematic uncertainty (see Supplementary Information). The side panels visualize the different character of the entanglement for attractive and repulsive interactions, where the ground state approaches two-body Bell states.}
        \label{fig:third}
\end{figure}

A quantitative measure of the strength of entanglement between the particles is given by the concurrence \cite{Wootters1998}. While its exact determination requires knowledge of the full density matrix, we can construct lower bounds as $C_1 = 2(\left| \rho_{2,3} \right| -  \sqrt{P_\textrm{LL} P_\textrm{RR}})$ and $C_2 = 2(\left| \rho_{1,4} \right| - \sqrt{P_\textrm{LR} P_\textrm{RL}})$ \cite{Jafarpour2012,Mazza,Kaufman2015}. Here, we have defined the density matrix in a position representation $|\alpha \beta\rangle$, where $\alpha$ and $\beta$ denote the spatial modes $\{\text{L},\text{R}\}$ of the $\ket\uparrow$ and $\ket\downarrow$ particle (Fig.~\ref{fig:third}a). For $C_1$ and $C_2$, we can extract the required quantities directly from the measured correlations \cite{Bonneau2018}: $\rho_{jj} \equiv P_{\alpha\beta}$ correspond to the populations of the spatial modes (Fig.~\ref{fig:third}b), while $\rho_{1,4},\rho_{2,3}$ are two-particle coherences which become apparent in the momentum correlations as oscillations along the relative and center-of-mass coordinates (Fig.~\ref{fig:third}c). A positive value of either $C_1$ or $C_2$ results in a concurrence $C(\rho)\geq \max(0, C_1, C_2) >0$, which demonstrates the presence of entanglement (see Fig.~\ref{fig:third}d).

Besides the entanglement between particles as measured by the concurrence, we are also interested in spatial mode entanglement between tweezer sites. Therefore, we study the entropy of entanglement \cite{Horodecki2009}, which compares the R\'{e}nyi or von Neumann entropy of generic partitions of a system to the entropy of the full system. If sub-systems individually have a higher entropy than the combined system, this proves the presence of entanglement between them \cite{Mintert2007}. We partition our system in spatial modes and spin modes (see Fig.~\ref{fig:first}c), which allows us to determine the role of either type of entanglement in the different interaction regimes. 

\begin{figure}[!tb]
				\includegraphics[width=\columnwidth]{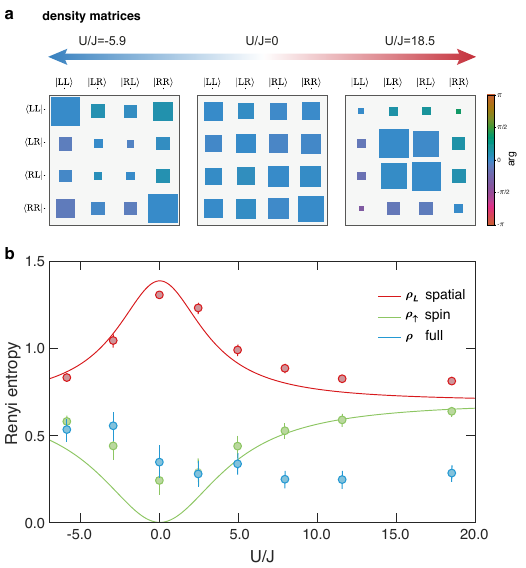}
        \caption{Entanglement entropy of the Hubbard dimer. a. To calculate entanglement entropy, we reconstruct the density matrix with a Bayesian quantum state estimation. The magnitude of the density matrix elements is proportional to the area of the squares, normalized to \num{0.5}. The phase is encoded in the color scale. b. Entanglement can occur between different subsystems, which may be taken to be the spatial or spin modes of the Hubbard dimer defined in Fig.\@ \ref{fig:first}c. Entanglement is present if the R\'{e}nyi entropy of a subsystem exceeds the R\'{e}nyi entropy of the full system. We observe entanglement of spatial modes for all interaction strengths and entanglement of spin modes at large interaction strengths. Error bars correspond to \SI{90}{\percent} credible intervals. The lines show the mode entropies of the ground state of the Hubbard dimer (see Supplementary Information).}
        \label{fig:fourth}
\end{figure}

We can evaluate the entanglement entropy of sub-systems directly from our correlation measurements (see Supplementary Information). The determination of the entropy of the full system, however, requires the knowledge of the full density matrix or collective measurements on multiple copies \cite{Islam2015}. To evaluate those density matrix elements which are not fully constrained by our correlation measurements (see, e.g., the sums in Fig.~\ref{fig:third}c), we implement a Bayesian estimate of the density matrix \cite{Blume-Kohout2010} such that it remains positive-semidefinite (see Fig.~\ref{fig:fourth}a and Methods). From this density matrix, we obtain a tightly constrained posterior distribution of the R\'{e}nyi entropy of the full system (blue circles in Fig.~\ref{fig:fourth}b). Comparing it to the entropy of the different subsystems, we see that the entanglement between spatial modes is largest for no interactions, while  entanglement between spin modes (which we identify with the particles) grows with increasing interaction strength \cite{Dowling2006}.

This disparate behavior of spatial and spin modes is an example of the inequivalence of entanglement between different degrees of freedom within the same state  \cite{Zanardi2002,Dowling2006,Tichy2011}. The entanglement between spin modes is the relevant quantity if the quantum state were to be used for quantum information processing or communication, where each party has control over exactly one spin sector. On the other hand, the entanglement entropy of spatial regions reflects the  cost of classically representing a quantum state with particle number fluctuations in real space. Our experiment realizes and probes the smallest non-trivial quantum systems in which these inequivalent notions of entanglement are both present.

These results show the potential of combined spin-resolved position and momentum correlation functions as a tool to characterize quantum states. The spin resolution and single-particle sensitivity of the detection method can be maintained for systems with larger particle number. It could be applied to continuum systems, for example to superfluid droplets or to measure order parameters in fermionic superfluids with nontrivial orbital symmetries \cite{Kitagawa2011}. Further applications extend to correlated few-body complexes such as individual, isolated Efimov trimers or fractional quantum Hall puddles, which could be mapped out completely in momentum space.

\paragraph*{Acknowledgements}
We gratefully acknowledge insightful discussions with Andrew Daley, Nicol\`{o} Defenu, Andreas Elben, Martin G\"{a}rttner, Philipp Hauke, and Marco Piani. This work has been supported by the ERC consolidator grant 725636, DFG grant JO970/1-1, the Heidelberg Center for Quantum Dynamics and is part of the DFG Collaborative Research Centre SFB 1225 (ISOQUANT). A.\@ B.\@ acknowledges funding from the International Max-Planck Research School (IMPRS-QD). P.M.P. acknowledges funding from European Unions Horizon 2020 programme under the Marie Sklodowska-Curie grant agreement No.\@ 706487 and from the Daimler and Benz Foundation.

A. B. and V. M. K. contributed equally to this work.





\balancecolsandclearpage


\setcounter{figure}{0}
\renewcommand{\figurename}{Supplementary Figure}

\renewcommand{\theequation}{S\arabic{equation}}

\newpage

\section*{Supplementary Information}

\paragraph*{Experimental sequence and parameters}
For our experiments we use two of the three lowest Zeeman sublevels of the hyperfine ground state of $^6$Li, labeled $\ket{1}$ and $\ket{3}$ in order of increasing energy. We realize a double-well potential with two optical tweezers of far-red-detuned laser light at \SI{1064}{\nm}. We generate and control the tweezers with an acousto-optical-deflector (AOD). This allows us to individually control each well of the double-well system. In order to prepare the system in its ground state, we deterministically prepare two atoms of different spin (labeled $\ket{\uparrow}\equiv \ket{1}$ and $\ket{\downarrow}\equiv \ket{3}$) in the ground state of a single tweezer \cite{deterministic}. Following the procedure described in \cite{Murmann2015}, we ramp on the second well and perform an adiabatic Landau-Zener passage to the ground state of the symmetric double-well. \NoteN{In a second ramp, we adiabatically increase the interaction strength by tuning the magnetic field and thereby changing the scattering length between the two atoms.} This system forms a Hubbard dimer, where we can tune the tunnel coupling $J$ by changing the global depth of the optical potential and the on-site interaction $U$ with a Feshbach resonance. \NoteN{We verify the adiabaticity of the ramps by reversing all the ramps and comparing the final population in the ground state of the single well with the initial one. Additionally, we observe that while the ramp to repulsive interaction is robust, at attractive interactions, the system becomes increasingly sensitive to residual tilts of the double-well potential. This leads to a significant occupation imbalance on the two sites for tilts smaller than $J$. Therefore, we measure only to moderate attractive interaction strengths.}

All presented measurements are performed in a double-well with a separation of $a= \SI{1.5}{\um}$ along the $x$-axis. Each tweezer has a waist of \SI{1.15}{\um} and single well trap frequencies of $\omega_{z} = 2 \pi \times \SI{3.95(10)}{\kHz}$ along the axial and $\omega_{x,y} = 2 \pi \times \SI{18.8(4)}{\kHz}$ along the in-plane directions, respectively. With this configuration, we achieve tunneling rates of $J/h = \SI{77(1)}{\Hz}$, where $h$ is Planck's constant.

In order to extract the in-situ populations and the momenta of the atoms, we employ a single-atom, spin-resolved imaging technique \cite{Bergschneider2018}.

\begin{figure}
\includegraphics[width=\columnwidth]{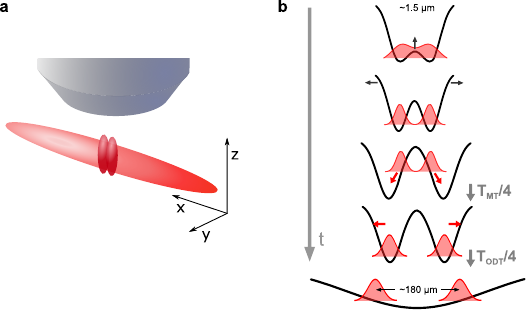}
\caption{\figfirst{Scheme for position-space measurements.} \sfigl{a} With the high-resolution objective, we create a double-well potential consisting of two adjacent optical tweezers. The double well is aligned along the long axis of the waveguide potential created by our optical dipole trap. \sfigl{b} In order to extract the in-situ populations, we suddenly increase the trap depth and separate the wells. After a $T_\text{MT}/4$ time evolution in the double-well potential, the atoms are released from the tweezers. An additional $T_\text{ODT}/4$ evolution in the optical waveguide potential increases the separation between the spatial modes by a factor of \num{120}. Note that the spatial dimensions in this figure are not to scale.}
\label{insitu}
\end{figure}

\paragraph*{Momentum measurements}
We measure the momentum of the atoms using a time-of-flight technique. After preparing the system, the quantum state is allowed to expand in a weak optical potential (optical dipole trap, ODT) which is elongated along the double-well axis with a longitudinal trap frequency of $\omega_{x} = 2\pi\times \SI{75}{\Hz}$ and transverse trap frequencies of $\omega_{y,z}= 2\pi\times \SI{600}{\Hz}$ (Supplementary Fig.~\ref{insitu}a). We image the atoms after a quarter of the axial trap period, $T_\textrm{ODT}/4=\frac{\pi}{2\omega_x}$. \Note{As the interactions during time of flight are negligible,} the unitary evolution in the dipole trap exactly performs a Fourier transform of the single-particle wavefunction in the $x$-direction. Neglecting the $y$ and $z$ coordinates, which are integrated out in the data analysis and imaging process, respectively, we obtain the initial momentum distribution along the double-well axis by a simple rescaling of the particle coordinates after time-of-flight, $k_{1,2}/k_\textrm{lat} = q x_{1,2}$, where $k_\textrm{lat}=\pi/a$ is the lattice momentum. We determine the scale factor from a fit as $q= \SI{20.1}{\per\mm}$, which is consistent with the trap frequency $\omega_x$ during time-of-flight. The spin-resolved correlation function $\langle n_\uparrow(k_1){n}_\downarrow(k_2)\rangle$ is obtained by averaging spin-resolved momentum measurements over several thousand runs. \Note{Note that the symmetry of the correlation functions (Fig.~2 and Supplementary Fig.~\ref{fulldata}) directly stems from the symmetries of the quantum state.}

\paragraph*{Position measurements}
The resolution of our imaging technique is limited to \SI{4}{\um} \cite{Bergschneider2018}. Therefore, we cannot directly measure the in-situ distribution of the double-well system. In order to reconstruct the position space distribution, we thus perform a three step scheme as illustrated in Supplementary Fig.~\ref{insitu}b). First, we project the wavefunction onto the individual wells by diabatically increasing the trap depth of the double-well to $\omega_{x,y} \approx 2\pi \times \SI{41.6}{\kHz}$ within \SI{2}{\ms}, decoupling the wells completely. Then we imprint a distinct and diametrical center of mass momentum onto the on-site wavefunctions of the two wells. We achieve this by a sudden change in the well separation to \SI{3.2}{\um} with a subsequent time evolution of around \SI{6}{\us} corresponding to a quarter of the on-site trap period $T_\text{MT}$. In the final step, we switch off the double-well potential followed by a time-of-flight evolution in the ODT (longitudinal trap frequency $\omega_x\approx 2\pi\times \SI{225}{\Hz}$). \Note{By tuning the trap depth of the individual wells, their final separation, and the trap frequency of the ODT, we can optimize the magnification of the on-site wavefunction and the magnification of the well separation independently.} For the parameters used in this paper, we separate the center of mass of the on-site wavefunction by approximately \SI{180}{\um} and achieve a fidelity for identifying each atom in the correct well of \SI{99.4(3)}{\percent}.
 
 \paragraph*{Data analysis and data set}
Each data set corresponds to $\unsim 4000$ momentum and $\unsim 1000$ position measurements. We postselect our data for images with exactly one atom per spin state, corresponding to $\geq \SI{80}{\percent}$ of all images. Before analysis, we group atom positions into \num{2}~pixel (or \SI{5.4}{\um}) bins. 

In total we measured the position and momentum correlation functions for \num{8} different values of $U/J$, ranging from attractive interaction ($U/J = -5.9$ at \SI{525}{\gauss}) to strong repulsive interaction ($U/J = 18.5$ at \SI{625}{\gauss}). Our full dataset is shown in Supplementary Fig.~\ref{fulldata}. All measurements are performed at a fixed tunnel coupling of $J/h = \SI{77(1)}{\Hz}$. The on-site interaction strength $U$ is set by the trap geometry and the s-wave scattering length $a_\textnormal{sc}$ which we set below a Feshbach resonance at $B=\SI{690}{\gauss}$. 
Both tunnel coupling and on-site interaction strength are calibrated as described in \cite{Murmann2015}.

\paragraph*{Hubbard dimer}

%
The ground state of the Hubbard dimer from Eq.\@  (1) can be written as
\begin{equation}
\psi_1 = \left( 1, \alpha_+(x), \alpha_+(x), 1 \right)/\sqrt{2(1+\alpha_+(x)^2)} \label{eq:eigenfun1}
\end{equation}
in a position space basis given by
\begin{eqnarray}
\ket{LL} &=& \cupleft \cdownleft \vac \nonumber \\
\ket{LR} &=& \cupleft \cdownright \vac \nonumber \\
\ket{RL} &=& \cupright \cdownleft \vac \nonumber \\
\ket{RR} &=& \cupright \cdownright\vac \text{,}
\label{basis}
\end{eqnarray}
%
with $x=U/4J$ and $\alpha_\pm(x) = x \pm \sqrt{1+x^2}$ \cite{Zanardi2002}.

A general (mixed) state takes the form
\beq
\bf{\rho} = 
 \begin{pmatrix}
  P_\textnormal{{LL}} & \rho_{{1,2}
} &  \rho_{{1,3}}&  \rho_{{1,4}} \\
 & P_\textnormal{{LR}} &  \rho_{{2,3}} &  \rho_{{2,4}}\\
    &   & P_\textnormal{{RL}} &  \rho_{{3,4}} \\
\textrm{h.c.}  & &  & P_\textnormal{{RR}}
 \end{pmatrix}\text{,}
 \label{rhomatrix}
\eeq
where $P_{\alpha \beta}$ are the populations and $\rho_{i,j}$ are the coherences.
 
\paragraph*{Evaluation of density matrix elements}

\Note{If we wish to determine the measured density matrix $\rho_\textnormal{exp}$ of a prepared state $\rho$, we can directly read off the populations from the spatial correlation measurements, while we can extract the coherences from the momentum correlation measurements as outlined in \cite{Bonneau2018,Klinkhamer2018} (Fig.~3a-c). To calculate the momentum correlation functions for $\rho$,} we use the Fourier transform of the single-particle basis states,
\begin{eqnarray}
\phi_\text{L} \equiv g(x+a/2) & \overset{FT}{\longrightarrow} & \tilde{g}(k) e^{-iak/2} \nonumber \\
\phi_\text{R} \equiv g(x-a/2) & \overset{FT}{\longrightarrow} & \tilde{g}(k)  e^{iak/2} \text{.}
\label{eq:fourierbasis}
\end{eqnarray}

Here, $g(x)$ is the on-site Wannier function, which is very well approximated by a Gaussian within the precision of the experiment. The momentum space representations of the single-particle modes share the envelope $\tilde{g}(k)$ (given by the Fourier transform of the on-site wavefunction $g(x)$) and differ only by a differential phase gradient $e^{iak}$.
In the basis defined by Eq.\@ (\ref{basis}) the momentum correlation operator $\hat{Z}= \hat{n}_\uparrow(k_1) \hat{n}_\downarrow(k_2)$ takes on the matrix representation 
\begin{equation}
\hat{Z} = \begin{pmatrix} 1 & e^{-iak_2} & e^{-iak_1} & e^{-ia(k_1+k_2)}\\ & 1 &e^{-ia(k_1-k_2)}&e^{-iak_1}\\ & & 1&e^{-iak_2}\\
   h.c.&&&1 \end{pmatrix},
\end{equation}
where we have neglected the envelope $\tilde{g}(k)$.

Using the momentum basis (\ref{eq:fourierbasis}), the expectation value of the operator $\hat{Z}$ can be calculated as
\begin{eqnarray}
\langle n_\uparrow(k_1){n}_\downarrow(k_2)\rangle& =& \Tr(\rho \hat{Z}) \nonumber\\
& =&  P_\text{LL}  + P_\text{LR}  + P_\text{RL}  + P_\text{RR}  \nonumber\\
& &+ 2\,\Re{\left\{\rho^{(1)} e^{i a k_1}\right\}} \nonumber\\ 
& &+ 2\,\Re{\left\{\rho^{(2)} e^{i a k_2}\right\}} \nonumber\\
& &+ 2\,\Re{\left\{\rho_{{2,3}} e^{i a( k_1-k_2)}\right\}} \nonumber\\ 
& &+ 2\,\Re{\left\{\rho_{{1,4}} e^{i a( k_1+k_2)}\right\}} ,
\label{densitymatrix}
\end{eqnarray}
with $\rho^{(1)}=\rho_{{1,3}}+\rho_{{2,4}}$, $\rho^{(2)}=\rho_{{1,2}}+\rho_{{3,4}}$ and $P_\text{LL}  + P_\text{LR}  + P_\text{RL}  + P_\text{RR} = \Tr(\rho) = 1$. 

For our data analysis, we use the quadrature representation
\begin{eqnarray}
\langle n_\uparrow(k_1){n}_\downarrow(k_2)\rangle & =& 1 \nonumber\\
& &+  2\,\Re{\left\{(\rho_{{1,3}}+\rho_{{2,4}})\right \}} \cos{a k_1} \nonumber\\ 
& & -  2\,\Im {\left \{(\rho_{{1,3}}+\rho_{{2,4}})\right \}} \sin{ a k_1} \nonumber\\ 
& &+  2\,\Re{\left\{(\rho_{{1,2}}+\rho_{{3,4}})\right \}} \cos{a k_2} \nonumber\\ 
& & -  2\,\Im {\left \{(\rho_{{1,2}}+\rho_{{3,4}})\right \}} \sin{ a k_2} \nonumber\\ 
& &+ 2\,\Re{\left\{\rho_{{2,3}} \right \} \cos{a( k_1-k_2)}} \nonumber\\
& &-2\,\Im{\left\{\rho_{{2,3}} \right \}} \sin{a( k_1-k_2)} \nonumber\\ 
& &+ 2\,\Re{\left\{\rho_{{1,4}} \right \} \cos{a( k_1+k_2)}} \nonumber\\
& &- 2\,\Im{\left\{\rho_{{1,4}} \right \} }\sin{a( k_1+k_2)},
\label{quadrature}
\end{eqnarray}
where the real and imaginary parts of the density matrix elements now explicitly appear as coefficients of the trigonometric basis functions of the momentum-space correlation function $\langle n_\uparrow(k_1){n}_\downarrow(k_2)\rangle$. 


\begin{figure}
	\includegraphics[width = \columnwidth]{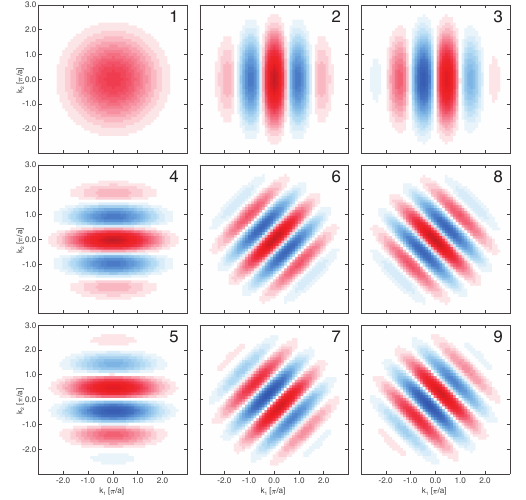}
	\caption{\figfirst{Basis functions $\textbf{B}_i$ used for the reconstruction of momentum correlations.} We obtain the envelope and fringe spacing from fits to single-particle density profiles after time-of-flight. The indices correspond to the lines in Eq.~\ref{quadrature}.}
	\label{basisfunctions}
\end{figure}

To obtain the off-diagonal matrix elements $\rho_{{1,4}}, \rho_{{2,3}}...$, we generate a reconstruction \textbf{R} of the measured momentum correlation functions \textbf{D} in terms of the known basis functions \textbf{B}, shown in Supplementary Fig.~\ref{basisfunctions}. We are looking for the weights $w$ such that the reconstruction  $\textbf{R} = \sum_i w_i \textbf{B}_i$ minimizes the total square error to the data, $\epsilon = \lbrack (\textbf{R}- \textbf{D}) | (\textbf{R}-\textbf{D})\rbrack$. Here quantities in bold are defined in the two-particle coordinate space spanned by $k_1$ and $k_2$ and $ \lbrack \cdot | \cdot \rbrack$ denotes bin-wise multiplication and summation over the entire space. Even though the basis functions $\textbf{B}_i$ are not perfectly orthogonal to each other \Note{due to the finite envelope function}, a closed form for the optimal weights can be given as $w_\textrm{opt}=Q^{-1} L$, where the matrix $Q$ quantifies the overlap of the basis functions via $Q_{ij}=\lbrack \textbf{B}_i | \textbf{B}_j\rbrack$ and $L$ is the overlap vector between the basis functions and the data, $L_i=\lbrack \textbf{B}_i | \textbf{D}\rbrack$. The optimal reconstruction for all data sets is shown in Supplementary Fig.~\ref{fulldata}. From the weights $w$ we read off the complex-valued off-diagonal density matrix elements according to Eq.~(\ref{quadrature}). The matrix elements $\rho_{{1,3}}$ and $\rho_{{2,4}}$ (as well as $\rho_{{1,2}}$ and $\rho_{{3,4}}$) contribute to the same features in the two-particle correlation functions and our measurements only reveal their complex sum.  Combining measurements from position and momentum space, we obtain 12 of the 16 real coefficients defining the density matrix. 

The dominant source of systematic errors on the optimal weights $w_\textrm{opt}$ are uncertainties in the fringe spacing as well as envelope waist and center in the basis functions \textbf{B}. These parameters as well as their uncertainties are obtained from a fit to the single-particle density distributions $\langle n(k) \rangle $ for the non-interacting data. We perform the reconstruction of $\langle n_\uparrow(k_1) {n}_\downarrow(k_2)\rangle$ 8000 times with basis function parameters randomly sampled from a normal distribution representing their uncertainty. The systematic error on the density matrix entries is given by their standard deviation over all instances of the basis function parameters.
 
We separately estimate the statistical error on the optimal weights $w_\textrm{opt}$ by resampling the measured two-particle probability distribution $\langle n_\uparrow(k_1) {n}_\downarrow(k_2)\rangle$ 1000 times with fixed basis function parameters, assuming independent shot noise in each bin. For each instance of the distribution, we obtain the reconstruction and the corresponding weights. The statistical error on the density matrix entries is given by the standard deviation of the distribution over all resampled instances. The reported errors on individual density matrix entries are quadrature sums of the systematic and statistical error bars.

We calculate the theory lines for the momentum density $\langle n(k_1) \rangle \equiv \langle n_\uparrow(k_1)\rangle + \langle {n}_\downarrow(k_1)\rangle$ and correlators $\chi(s)$ and $\xi(d)$ by performing the corresponding integrals over the reconstruction $\textbf{R}$. 

\paragraph*{Entanglement witness from correlators} A very direct way to witness entanglement between spin modes is provided by analysing the correlators $\chi(s)$ and $\xi(d)$ and comparing their amplitudes to limits compatible with separable states. From Eq.~(\ref{densitymatrix}), the single-particle densities are given by $n_\uparrow(k_1) = 1+2\,\Re{\left\{\rho^{(1)} e^{i a k_1}\right\}}$ and $n_\downarrow(k_2) = 1+2\,\Re{\left\{\rho^{(2)} e^{i a k_2}\right\}}$. The correlators take the form 
\beq
\xi(d) = \frac{1+2 |\rho_{2,3}| \cos{(ad + \phi_{2,3})}}{1+ 2 |\rho^{(1)}||\rho^{(2)}|   \cos{(ad - \phi^{(1)} + \phi^{(2)})}},
\eeq
where we have written the complex density matrix elements in polar representation, $\rho_j=|\rho_j|e^{i \phi_j}$.

We follow \cite{Kaufman2015} to find the maximum contrast in $\xi$ . For a single product state, i.e., $\rho^\textrm{(prod)} = \rho^{(\uparrow)} \otimes \rho^{(\downarrow)}$ , triangle inequalities on the density matrix imply that 

\begin{align}
\abs{\rho^\textrm{(prod)}_{2,3}} &\leq \abs{\rho^{(\uparrow)}_\text{1,2}} \abs{\rho^{(\downarrow)}_\text{2,1}}\nonumber\\
&\leq\sqrt{\rho^{(\uparrow)}_\text{1,1} \rho^{(\uparrow)}_\text{2,2}} \sqrt{\rho^{(\downarrow)}_\text{1,1} \rho^{(\downarrow)}_\text{2,2}}\nonumber\\
&=\sqrt{P_\textrm{RR} P_\textrm{LL}} = \sqrt{P_\textrm{LR} P_\textrm{RL}} \text{,}
\label{eq:bound23}
\end{align}
where we choose the tighter bound $\sqrt{P_\textrm{RR} P_\textrm{LL}}$.

For the most general form of a separable state, $\rho^\textrm{(mix)} = \sum_{i} \lambda_{i} \rho^{(i,\uparrow)} \otimes \rho^{(i,\downarrow)}$, one still finds that 
\begin{align}
\abs{\rho^\textrm{(mix)}_{2,3}} \leq \sqrt{P_\textrm{RR} P_\textrm{LL}}.
\end{align}
The in-situ occupation probabilities hence set an upper bound on the modulus of $\rho_{2,3}$ \cite{Kaufman2015}.

The strongest correlations that are compatible with a separable state are then given by 
\begin{align}
\xi_\textrm{max} & = \frac{1+ 2 \sqrt{P_\textrm{RR} P_\textrm{LL}}}{1-2 |\rho^{(1)}|\,|\rho^{(2)}|} \nonumber\\
\xi_\textrm{min} & = \frac{1- 2 \sqrt{P_\textrm{RR} P_\textrm{LL}}}{1+2 |\rho^{(1)}|\,|\rho^{(2)}|} \text{,}
\end{align}
with similar expressions for $\chi(s)$. The upper and lower bounds on the correlators under the assumption of separability are shown as the gray shaded area in Fig.~2c.

\paragraph*{Concurrence}
We can use a lower bound of the concurrence $C$ to quantify the amount of entanglement in our system. For a pair of two-level systems (qubits), the concurrence can be used to obtain the entanglement of formation. In the case of a pure two-qubit state $\Phi$, $C$ is formally defined as $C(\Phi) = |\bra{\Phi}(\sigma_y \otimes \sigma_y)\ket{\Phi^*}|$ \cite{wootters} with the Pauli matrix $\sigma_y = \begin{pmatrix} 0 & -i \\ i&0 \end{pmatrix}$. For a mixed state, one defines the concurrence as the infimum of its value over all pure state decompositions
\begin{equation}
C(\rho) = \inf \sum_i {p_i C(\Phi_i)}.
\end{equation}

In the special case of a pair of qubits one can find the explicit formula
\begin{equation}
C(\rho) = \max\{0,\lambda_1 - \lambda_2 - \lambda_3 -\lambda_4\}
\end{equation}
where $\lambda_i^2$, $i\in\{1,2,3,4\}$, are the eigenvalues of $\rho(\sigma_y \otimes \sigma_y)\rho^*(\sigma_y \otimes \sigma_y)$ in decreasing order. $C(\rho)$ can take values from $0$ for a product state to $1$ for a maximally entangled state. A positive value of the concurrence indicates the presence of entanglement in the system. In particular, for a two-qubit system, the entanglement of formation can directly be calculated from the concurrence \cite{wootters}.

For our purposes, we follow \cite{Jafarpour2012} to construct a lower bound for the concurrence and hence for the entanglement of formation as: 

\begin{equation}
C(\rho) \geq \max\{0,2(|\rho_\text{1,4}| - \sqrt{P_{LR}P_{RL}}),2(|\rho_\text{2,3}| - \sqrt{P_{LL}P_{RR}})\}.
\end{equation}
Hence, from the matrix elements obtained from position and momentum space correlations, we can directly calculate the lower bounds of the concurrence (see Fig.~3d).

\paragraph*{Measurement of the full density matrix}
Complete knowledge of the density matrix can in principle be obtained by performing rotations on the state prior to measurement. Reference \cite{Bonneau2018} suggests to apply pulses of pure tunneling or tilt to the double-well. We point out that it is also possible to use a pulse with interaction only, i.e. to apply Hamiltonian (1) with $J=0$ for a time $t=\frac{1}{4}\frac{h}{U}$, which allows the measurement of the complex differences $\rho_{{1,3}} - \rho_{{2,4}}$ etc. and completes the measurement of all entries of the density matrix.

\Note{Alternatively, correlated position-momentum measurements would deliver the density matrix elements complementary to the entries obtained from pure position and momentum correlation measurements \cite{Taguchi2008}. These measurements could, for example, be performed in an intermediate basis.}

\paragraph*{Reconstruction of the density matrix}

\begin{figure}
	\includegraphics[width = \columnwidth]{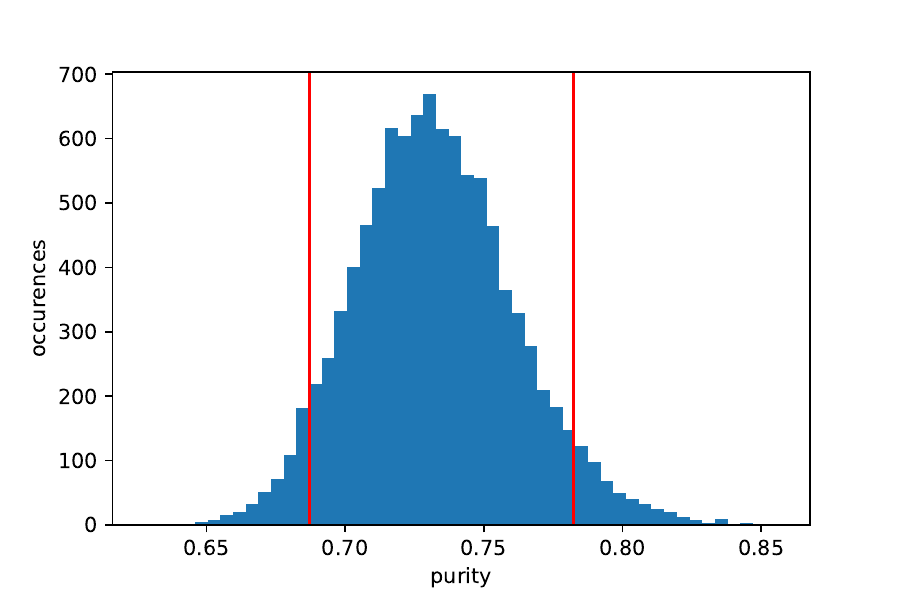}
	\caption{\figfirst{Posterior distribution of the full system purity $V$.} The reconstructed density matrix is obtained from measurements performed at $U/J = 18.5$. The red lines mark the \SI{90}{\percent} credible interval.}
	\label{hist_purity}
\end{figure}

\Note{From the previously described methods, we obtain only a subset of the parameters required to unambiguously describe the density matrix of the prepared state. This means that there is a set of density matrices which would be consistent with the measured parameters. Also, we determine these parameters with statistical and systematic uncertainties, which can lead to unphysical properties for an entire set of density matrices consistent with $\rho_{\textrm{exp}}$. For example, their eigenvalues with small magnitudes may consistently turn out to be negative.}

\begin{figure*}[!ht]
	\includegraphics[width = 0.8\textwidth]{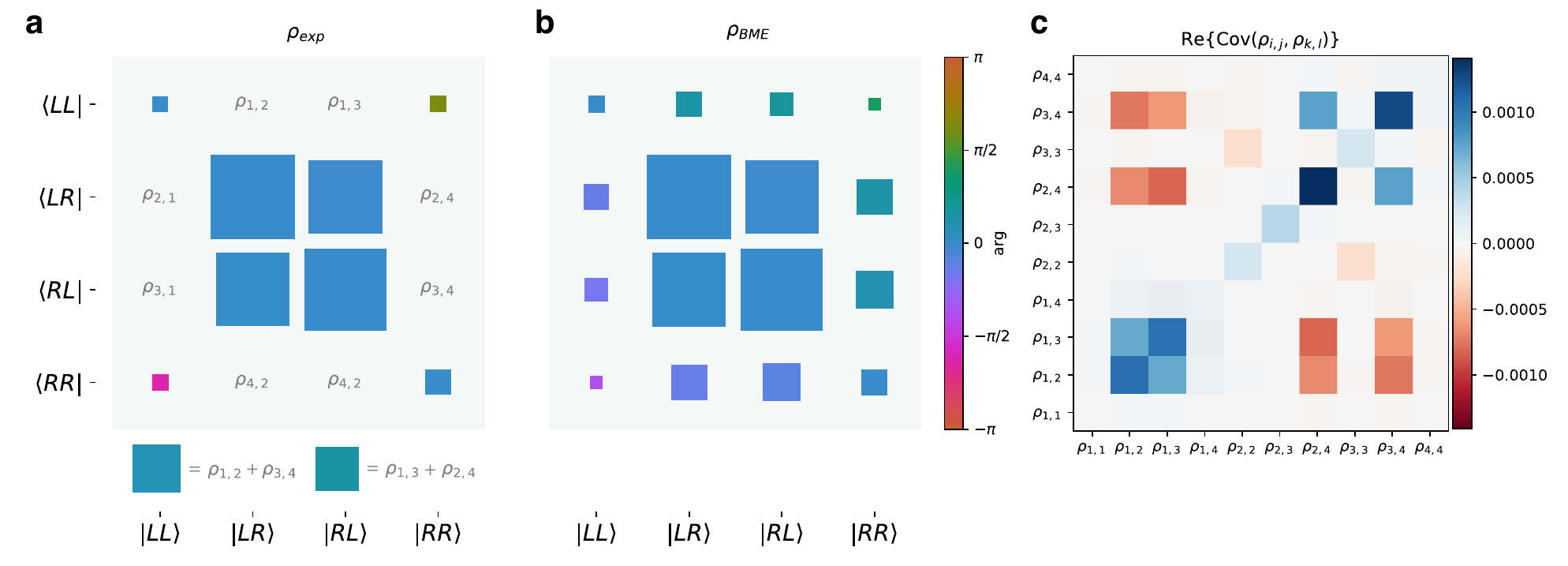}
	\caption{\figfirst{Density matrix reconstruction.} \sfigl{a} Measured density matrix at $U/J = 18.5$, obtained from the spatial measurements and from fitting the momentum correlations. The magnitude of the entries is proportional to the area of the squares and normalized to \num{0.5}, while the phase is indicated by the color scale. \sfigl{b} Reconstructed density matrix obtained from the Bayesian mean estimate. \sfigl{c} Real part of the covariance of the reconstructed density matrix elements obtained from sampling. While the diagonal elements are correlated by the condition $\Tr\rho = 1$, some off-diagonal elements are anti-correlated due to, e.g., $\rho^{(1)} = \rho_{{1,3}}+\rho_{{2,4}}$.}
	\label{rhocovariance}
\end{figure*}

To avoid these unphysical sets of density matrices, we reconstruct the density matrix $\rho_{\textrm{BME}}$ using a Bayesian quantum state estimation as outlined in \cite{Blume-Kohout2010}.
All information about the experiment is contained in the likelihood function $\mathcal{L}(\rho) = p(\mathcal{M}|\rho)/p(\mathcal{M})$, a distribution over the measured data $\mathcal{M}$ conditioned on a certain hypothesis about the state $\rho$. 
It quantifies the relative plausibility of different possible states. 
In the case of Gaussian distributed errors, it takes the form
\begin{equation}
 \mathcal{L}(\rho) =   \prod_j\frac{1}{\sqrt{2\pi\sigma_j^2}} \exp{ \left(\frac{-(\mathcal{M}_j- Tr[\hat{\mathcal{M}}_j \rho])^2}{2\sigma_j^2} \right)}
\end{equation}
with the set of measurements $\mathcal{M} = \{\mathcal{M}_j\}$ and corresponding errors $\sigma_j$. It contains the four real-valued populations and the complex coherences defined as coefficients in Eq.~\ref{quadrature}. The corresponding operators $\hat{\mathcal{M}}_j$ denote the projections of the state onto these entries.
To obtain the posterior distribution $\pi_f(\rho)d\rho$, the likelihood is multiplied with a prior distribution $\pi_0(\rho)d\rho$ on the states:
\begin{equation}
\pi_f(\rho)d\rho \propto \mathcal{L}(\rho) \pi_0(\rho)d\rho
\end{equation}
 where the proportionality is up to normalization.
Because no prior knowledge on the states is assumed, we choose the Hilbert-Schmidt prior as an uninformative prior over all density matrices that meet the requirements of positive definiteness and unity trace.
The Bayesian mean estimate $\hat{\rho}_{\textrm{BME}}$ is then given by the mean of the posterior distribution
\begin{equation}
\hat{\rho}_{\textrm{BME}} = \int \rho \pi_f(\rho)d\rho.
\end{equation}
The expectation value of an observable $\mathcal{O}$ can be calculated as $\langle\mathcal{O}\rangle = \int \mathcal{O}(\rho) \pi_f(\rho)d\rho$ with the errors given in terms of the credible interval of its posterior distribution.
An example of this can be seen in Supplementary Fig.~\ref{hist_purity} for the purity.

To compute $\hat{\rho}_{\textrm{BME}}$, we first parametrize the density matrix as $\rho = \hat{T}^\dagger \hat{T}$ where $\hat{T}$ is a random complex matrix with \num{32} real parameters $\vec{t}$.
This form ensures that $\rho(\vec{t})$ is a positive semidefinite and Hermitian matrix of trace one.
We sample the posterior using Hamiltonian Monte Carlo (HMC) to compute $\hat{\rho}_{\textrm{BME}}$ and the subsequent entanglement measures and determine their uncertainty.

In the experiment, not all entries of $\rho_\text{exp}$ are measured individually. For $\rho_{1,2}, \rho_{2,4}$ and $\rho_{2,3}, \rho_{3,4}$  only the sums are determined.
This can be incorporated naturally in the Bayesian estimation by specifying the set of measurements $\mathcal{M}$ accordingly.
The HMC procedure then samples the space of physically possible entries while leaving the corresponding sum unchanged.
The uncertainty about the entanglement measures with respect to the exact distribution of the sum constituents is therefore expressed in the credible intervals of these values.

The covariances of $\rho_{i,j}$ shown in Supplementary Fig.~\ref{rhocovariance} support this intuition, as the variance is largest for $\rho_{2,4}$ (and  $\rho_{3,4}$) while being most anticorrelated with $\rho_{1,2}$ ($\rho_{2,3}$ respectively).

\paragraph*{R\'{e}nyi entropy}
With our Bayesian estimate of the density matrix, we can evaluate the expected value of the R\'{e}nyi entropy $S = - \log \Tr(\rho^2)$ for the entire system and for different sub-systems. In our definition, we use the natural logarithm, and $\Tr(\rho^2) \equiv V$ can be identified as the purity.

The entanglement between the spin modes is obtained by tracing out one of the two particles. In term of the full-system density matrix in Eq.~(\ref{rhomatrix}), the single-spin density matrix is given by

\beq
\rho_\uparrow = \begin{pmatrix}
P_{LL} + P_{LR} & \rho_{1,3} + \rho_{2,4} \\
\textrm{h.c.}&P_{RL} +P_{RR} & \\
\end{pmatrix}
\eeq
with the single-spin R\'{e}nyi entropy $S_\uparrow = -\log{\Tr{(\rho_\uparrow^2)}}$. Note that the single-spin entropy depends only on the sum of the matrix elements $ \rho_{1,3} + \rho_{2,4}$ as well as the populations and can be extracted directly from our measurements.

The single-site density matrix in the basis $\ket{\uparrow \downarrow}$, $\ket{\uparrow}$, $\ket{\downarrow}$, $\vac$ is 

\beq
\rho_L
=
\begin{pmatrix}
    P_{LL} &  & & \\
         & P_{LR}  & & \\
             &  &P_{RL} & \\
                 &  & & P_{RR}\\
\end{pmatrix}
\eeq
i.e. it has only the populations, but no coherences and we calculate its entropy as $S_\textrm{L} = -\log{\Tr{(\rho_L^2)}}$.

In Fig.~4b, we compare the measured entanglement entropies to the Hubbard model using the density matrix from Eq.~(\ref{eq:eigenfun1}).

\balancecolsandclearpage
\clearpage

\begin{figure*}
	\includegraphics[angle=90,origin=c,width=1.0\textwidth]{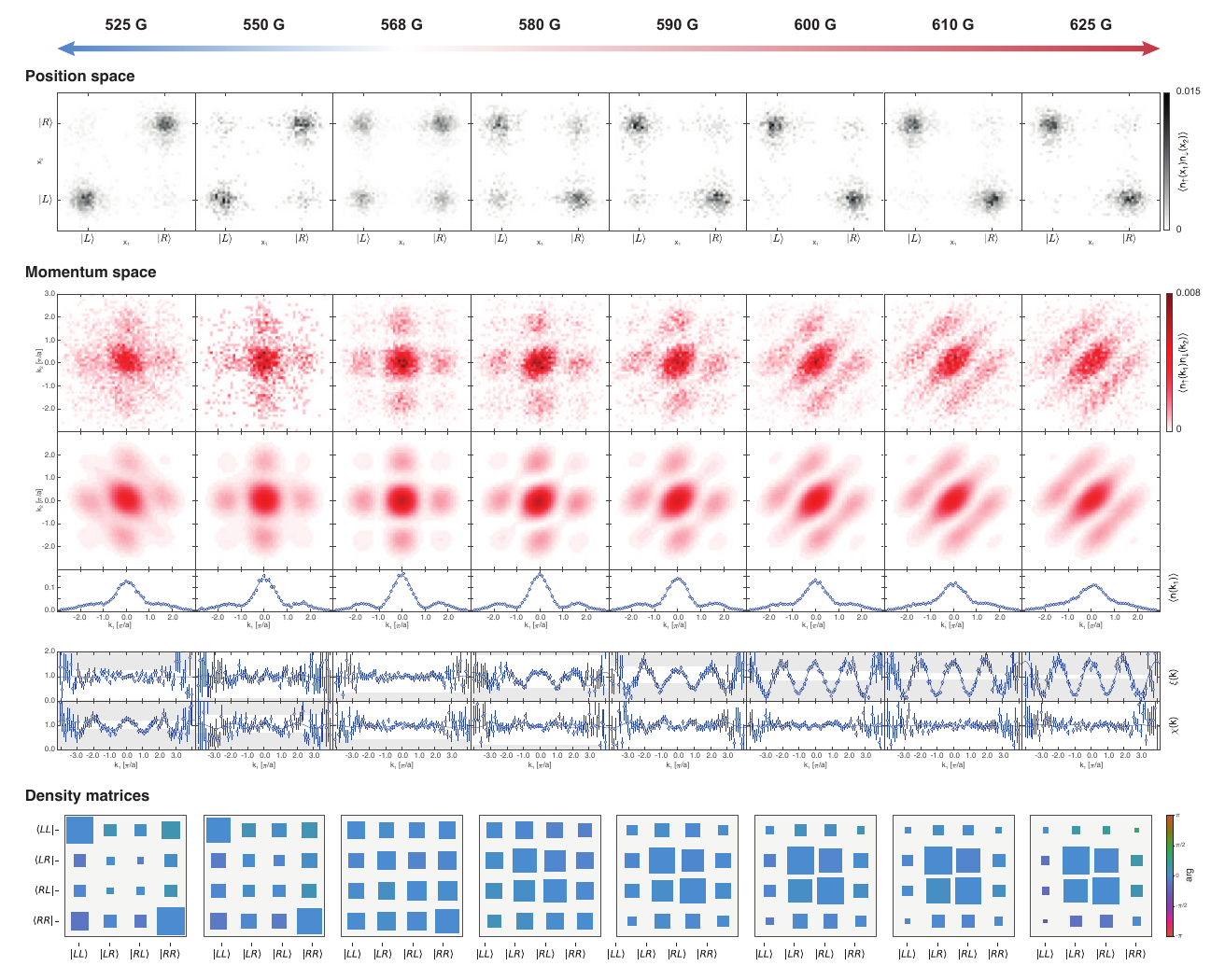}
	\caption{\figfirst{Full data set.} Insitu (top) and momentum space correlations (middle) for all measured interaction strengths. From left to right, the magnetic offset field tunes the interactions from attractive (\SI{525}{G}) through zero (\SI{568}{G}) to the strongly repulsive regime (\SI{625}{G}). The bottom row shows the reconstructed density matrices $\hat{\rho}_{\textrm{BME}}$.}
	\label{fulldata}
\end{figure*}

\end{document}